\documentclass{appolb}
\usepackage{epsfig}

\begin{document}

\title{\Large\bf Kinetics of growth process controlled by mass-convective fluctuations and finite-size curvature effects}
\author{A. Gadomski, J. Si\'odmiak\address{Department of Theoretical Physics, Institute of Mathematics and Physics,\\
University of Technology and Agriculture,\\Kaliskiego 7, Bydgoszcz
PL--85796,  Poland} \and I.Santamar\'ia-Holek\address{Facultad de
Ciencias, Universidad Nacional Autonoma de Mexico.\\ Circuito
exterior de Ciudad Universitaria, 04510, D. F., Mexico}
\and J. M. Rub\'i\address{Departament de F\'isica Fonamental,\\
E-08028 Barcelona, Spain} \and M. Ausloos\address{University of
Li$\grave{e}$ge, SUPRATECS, Li$\grave{e}$ge B--4000, Euroland}}
\date{}
\maketitle

%\newpage
\begin{abstract}
In this study, a comprehensive view of a model crystal formation
in a complex fluctuating medium is presented. The model
incorporates Gaussian curvature effects at the crystal boundary as
well as the possibility for superdiffusive motion near the crystal
surface. A special emphasis is put on the finite-size effect of
the building blocks (macroions, or the aggregates of macroions)
constituting the crystal. From it an integrated static-dynamic
picture of the crystal formation in terms of mesoscopic
nonequilibrium thermodynamics (MNET), and with inclusion of the
physically sound effects mentioned, emerges. Its quantitative
measure appears to be the overall diffusion function of the
formation which contains both finite-size curvature-inducing
effects as well as a time-dependent superdiffusive part. A quite
remarkable agreement with experiments, mostly those concerning
investigations of dynamic growth layer of  (poly)crystaline
aggregation, exemplified by non-Kossel crystals and biomolecular
spherulites, has been achieved.
\end{abstract}


\section{Introduction}

The present study is devoted to modeling a formation of a soft
(ordered) material known as the non--Kossel crystal
\cite{AGJS_PSSB}, as well as to propose how to model a
polycrystalline aggregate termed the soft-matter spherulite
\cite{AGJL_IJQCh}. The approach we are exploring throughout this
paper assumes that the system of interest is a mesoscopic system
\cite{DREGCO}, by its nature combining both classical and
quantum-mechanical properties, though we are offering here a
description based on nonequilibrium statistical thermodynamics,
taken suitably at a mesoscopic (molecular cluster) level
\cite{ISHth}, i.e. a classical limit of the approach can
preferentially be exploited \cite{JMR1}.

In general, examples of soft materials include biopolymers and
charged polymer solutions, protein/membrane complexes, colloidal
suspensions, gels, nucleic acids and their assemblies, to mention
but a few. Most forms of condensed matter - except of metals and ceramics (perhaps) \cite{ausl} - are
soft and these substances are composed of aggregates and
macromolecules, with interactions that are too weak and complex to
form crystals spontaneously; therefore, the task of how to grow
crystals in complex environments still remains a real challenge
\cite{CHERN1}.

Thus, the phrase 'soft condensed matter' has been coined, also for
the clear reasons mentioned below. First, a striking feature is
that slight perturbations in temperature, pressure or
concentration, or small (piconewton) external forces, can all be
enough to essentially induce microstructural changes. Second,
thermal fluctuations are almost by definition pronounced in soft
materials, and in consequence, entropy is a privileged determinant
of a soft microstructure formation, so that disorder, slow
dynamics and kinetics, and plastic deformation are the rules
\cite{dGM}. Although soft materials have attracted engineers for
ages, only recently have physicists taken an interest in such
materials, and attempted to implement what is the essence of
physics, that is to produce, at various levels of description,
simple models that include the possible minimum information
required to explain relevant features, cf. an example concerning
proteins \cite{DIMAT}.

Nowadays, especially with the advent of single molecule
spectroscopies are we managing to get quantitative data on things
such as aggregation (and folding) effects in complex biomolecular
environments, motion of molecular motors \cite{REIMAN}, electron
and proton transfer rates \cite{Ewa_GN}, etc., that are
sufficiently reliable to formulate and test simple models. The
availability of new experimental tools and simulation capability
are urging physicists to apply their methods to structural
biology, i.e. to allow the prediction of structure on the basis of
known microscopic forces.

A key issue is the fact that the environments in which a formation
of the soft material happens are characteristic of pronounced
spatial variations in dielectric constant, e.g. water and lipids
\cite{AGJS2}. It is of major importance to realize that
competition between interactions of different ranges results in
different types of aggregation of molecules. The starting point
should thus be about a discussion of the relative role of the
various fundamental interactions in such systems (electrostatic,
hydrophobic, conformational, steric, van der Waals, etc.), and
what is their real impact on the aggregation of possibly ordered
(crystal formation) and/or disordered types. The next focus could
be on how these competing interactions influence the form (and,
topology) of soft and biological matter, like biopolymers and
proteins, leading to self-assembling systems of the type listed
above \cite{PHYSBIOMATR}.

In the underlying study, let us propose a way of modeling the soft
matter objects named non--Kossel crystals \cite{CHERN1}, and in
part also the biopolymeric spheroidal polycrystals, commonly
termed spherulites. The way that we have chosen to achieve the
above stated goal is based on the assumption that the biomolecular
system, generally out of equilibrium, can be best described at a
mesoscopic level, using the conception of mesoscopic
nonequilibrium thermodynamics (MNET) \cite{JMR}, section 2. In
sections 3 and 4, we will offer the thermodynamic-kinetic
description of the soft material formation first by introducing
its deterministic part (section 3), and second, by going to its
stochastic, no doubt, more interesting part (section 4). In the
deterministic part, we wish to place our emphasis on an
unquestionable basis of each soft material formation, namely, how
does the material formation depend upon the boundary effect, being
typically curvature-dependent, and having also - in the case of
spherulites - a nonequilibrium kinetic account readily involved in
the overall process. In the stochastic part, in turn, we are lucky
to support our mesoscopic view by a microscopic picture of a
sub-process, essentially limiting the model material formation in
the crystal boundary layer \cite{vekir}. Here, we have in mind the
physical fact that the growth of the objects that we model can
thoroughly be controlled by the macroion velocity field of the
crystal ambient phase nearby its interface with the growing
crystal. Since it may lead to some very interesting physical
consequences, e.g. a time-dependent viscosity effect \cite{plonka}
(section 4), we are going to exploit this experimentally justified
observation \cite{EXPPROT} in a sufficient detail, looking at the
macroion finite-size effect (FSE)\footnote{In our approach FSE
readily assigned to a macroion, will be emphasized twofold. First,
it enters the nonequilibrium boundary condition of modified
Gibbs--Thomson type (section 3). Second, it is involved in the
change of viscosity near the crystal vs surroundings interface,
the so-called size-dependent viscosity effect, cf. section 4.
Thus, the FSE is proposed to be a bridge between stochastic and
deterministic parts of the approach. It will, for example enter
the overall diffusion function of the complex crystal formation,
$D(R,t)$, see section 4.}, and its impact on the overall
crystallization process seen as a combination of static
(curvature-oriented) and dynamic (constrained motion-oriented)
effects in the crystal's growth layer \cite{vekir}.
%(cooperative) diffusion function of the crystallizing medium, as
%well as on some examination of the correlational macroion velocity
%field mostly in the time domain.
Concluding address, contained in
section 5, closes the paper.

\section{MNET and its application to nucleation and growth phenomena}

Mesoscopic nonequilibrium thermodynamics (MNET) provides a general
framework from which one can study the dynamics of systems defined
at the mesoscale \cite{JMR,JmrAg}, and has been applied to analyze
different irreversible processes taking place at those scales
\cite{ISHth}. The formulation of the theory is based on the fact
that a reduction of the observational time and length scales of a
system usually entails an increase in the number of degrees of
freedom which have not yet equilibrated and that therefore exert
an influence on the overall dynamics of the system. Those degrees
of freedom $\gamma$ ($\equiv\{\gamma_{i}\}$) may for example
represent the velocity of a particle, the orientation of a spin,
the size of a macromolecule or any coordinate or order parameter
whose values properly define the state of a mesoscopic system in a
phase  space.  The characterization of the state of the system
essentially relies on
%precises of
the knowledge of $P(\gamma,t),$   the probability
density of finding the system at the state
$\gamma\in(\gamma,\gamma+d\gamma)$ at time  $t.$  One can then
formulate the Gibbs entropy postulate \cite{dGM,vKampen} in the
form
\begin{equation}
S=S_{eq}-k_{B}\int
P(\gamma,t)\ln\frac{P(\gamma,t)}{P_{eq}(\gamma)}d\gamma\; .
\label{entropy postulate}
\end{equation}
Here $S_{eq}$ is the entropy of the
system when the degrees of freedom $\gamma$ are at equilibrium. If
they are not, the contribution to the entropy arises from
deviations of the probability density $P(\gamma,t)$ from its
equilibrium value $P_{eq}(\gamma)$ given by
\begin{equation}
P_{eq}(\gamma )\sim\exp\left(\frac{-\Delta{\cal
W}(\gamma)}{k_{B}T}\right), \label{peq general}
\end{equation}
where $\Delta{\cal W}\equiv\Delta{\cal W}(\gamma)$ is the minimum
reversible work required to create that state \cite{Landau},
$k_{B}$ is Boltzmann's constant, and $T$ is the temperature of the
heat bath. Variations of the minimum work for a thermodynamic
system are expressed as

\begin{equation}
\Delta{\cal W}=\Delta E-T\Delta S+p\Delta V-\mu\Delta
M+\sigma\Delta \Sigma+\ldots\,, \label{minimum work}
\end{equation}
where (using a standard notation) extensive quantities refer to
the system and intensive ones to the bath. The last term
represents the work performed on the system to modify its surface
$\Sigma$, whereas $\sigma$ stands for the surface tension.

To obtain the dynamics of the mesoscopic degrees of freedom one
first takes variations in Eq. (\ref{entropy postulate})
\begin{equation} \delta S=-k_{B}\int\delta
P(\gamma,t)\ln\frac{P(\gamma,t)}{P_{eq}(\gamma)}\, d\gamma.
\label{variations of Gibbs entropy post}
\end{equation}
focusing only on the nonequilibrated degrees of freedom.

The probability density evolves in the $\gamma-$space according
with the continuity equation
\begin{equation}
\frac{\partial P(\gamma,t)}{\partial t}=-\frac{\partial
J(\gamma,t)}{\partial\gamma},
\label{continuity1}
\end{equation}
where $J(\gamma,t)$ is an unknown probability current. To obtain
its value, one proceeds to derive the expression of the entropy
change, $dS/dt$, which follows from the continuity equation
(\ref{continuity1}) and the Gibbs equation (\ref{variations of
Gibbs entropy post}). After a partial integration, one then
arrives at

\begin{equation}
\frac{dS}{dt}=-\int\frac{\partial}{\partial\gamma}J_{S}d\gamma+\sigma_e,
\label{entropychange}
\end{equation}
where $J_{S}=J(\gamma, t)ln\frac{P(\gamma, t)}{P_{eq}(\gamma)}$ is
the entropy flux, and
\begin{equation}
\sigma_e=-k_{B}\,\int
J(\gamma,t)\,\frac{\partial}{\partial\gamma}\left(\ln\frac{P(\gamma,t)}{P_{eq}(\gamma)}\right)\,
d\gamma, \label{sigma1}
\end{equation}
is the entropy production which is expressed in terms of currents
and conjugated thermodynamic forces defined in the space of
mesoscopic variables. We will now assume a linear dependency
between fluxes and forces and establish a linear relationship
between them
\begin{equation}
J(\gamma,t)=-k_{B}L(\gamma)\,\frac{\partial}{\partial\gamma}\left(\ln\frac{P(\gamma,t)}{P_{eq}(\gamma)}\right),
\label{corriente1}
\end{equation}
 where $L\equiv L(\gamma)$ is an Onsager coefficient, which depends on the
mesoscopic coordinates $\gamma$; in general, it also depends on
the state variable $P(\gamma,t)$. To derive this expression,
locality in $\gamma-$space has also been taken into account, for
which only fluxes and forces with the same value of $\gamma$
become coupled.

The resulting kinetic equation then follows by substituting Eq.
(\ref{corriente1}) back into the continuity equation
(\ref{continuity1})

\begin{equation}
\frac{\partial P(\gamma, t)}{\partial
t}=\frac{\partial}{\partial\gamma}\left(D(\gamma, t)
P(\gamma,t)\frac{\partial}{\partial\gamma}ln\frac{P(\gamma,
t)}{P_{eq}(\gamma)}\right),\label{F-P}\end{equation} where we have
defined the diffusion coefficient as $D(\gamma,
t)\equiv\frac{k_{B}L(\gamma)}{P(\gamma, t)}$. This equation, which
in view of Eq. (\ref{peq general}) can also be written as

\begin{equation}
\frac{\partial P(\gamma, t)}{\partial
t}=\frac{\partial}{\partial\gamma}\left(D(\gamma, t)\frac{\partial
P(\gamma, t)}{\partial\gamma}+\frac{D(\gamma,
t)}{k_{B}T}\frac{\partial\Delta{\cal W}}{\partial\gamma}P(\gamma,
t)\right)\,,\label{F-P drif diffusion}\end{equation} is the
Fokker-Planck equation accounting for the evolution of the
probability density in $\gamma$-space.

\subsection{Nucleation processes}

The expression of the nucleation rates can be obtained from the
proposed formalism \cite{JMR1}. To this purpose one has to
interpret the nucleation (or in general any activated process) as
a diffusion process along the mesoscopic coordinate describing the
state of the system, the embryos, at short time scales in between
the metastable and the crystal phases. The reaction rate or
diffusion current can be written in terms of the fugacity
$z\equiv\exp\mu/k_{B}T$ as
\begin{equation}
\label{fug1}
J(\gamma, t)=-k_{B}L\frac{1}{z}\frac{\partial
z}{\partial\gamma}.
\end{equation}
The current can also be expressed as
\begin{equation}
\label{fug2}
J(\gamma, t)=-D(\gamma, t)\frac{\partial
z}{\partial\gamma},
\end{equation}
where $D(\gamma, t)=k_{B}L/z$ represents the diffusion
coefficient. As a first approximation we assume that $D(\gamma,
t)=D=const.$ and integrate from the initial to the final position,
obtaining
\begin{equation}
\overline{J}\equiv\int_{\mu_1}^{\mu_2}Jd\gamma=-D(z_{2}-z_{1})=-D(\exp\frac{\mu_{2}}{k_{B}T}-\exp\frac{\mu_{1}}{k_{B}T}).
\end{equation}
This equation can alternatively be expressed as
\begin{equation}
\overline{J}=J_{0}\left(1-e^{A/k_{B}T}\right),
\end{equation}
where $\overline{J}$ is the integrated rate,
$J_{0}=Dexp(\mu_{1}/k_{B}T)$ and $A=\mu_{1}-\mu_{2}$ is the
affinity. We have then shown that a mesoscopic thermodynamic
analysis may lead to the formulation of the nonlinear kinetic laws
governing nucleation processes. Nucleation processes in which the
embryos are embedded in an inhomogeneous bath can also be studied
by means of the theory proposed \cite{JMR1}.

\subsection{Agglomeration and growth processes}
MNET can in general be used to study kinetic processes taking
place in mesostructures such as the nanostructure arrays
\cite{bimberg}. The size of these structures is in between those
of single particles and macroscopic objects. They carry out
assembling (clustering) and impingement (pattern formation)
processes. They may also diffuse in a thermal bath, be convected
by external flow and be affected by external driving forces as
those acting in extrusion, shearing, injection processes involved
in the mechanical processing of a melt. The growth rate of the
agglomerates can be determined by the formalism previously
presented in which the mesoscopic variable is the volume of the
grain. One formulates the Gibbs entropy postulate in terms of the
probability density $P(\mathrm{v},t)$, where now the degree of
freedom $\gamma$ is $\mathrm{v}$ - the volume of the grain or
molecular cluster. Proceeding as indicated previously one obtains
the growth rate
\begin{equation}
J(\mathrm{v},
t)=\frac{L(\mathrm{v})}{TP(\mathrm{v},t)}\left[k_BT\frac{\partial
P(\mathrm{v}, t)}{\partial \mathrm{v}}+P(\mathrm{v},
t)\frac{\partial \Phi}{\partial \mathrm{v}}\right]
\end{equation}
Interpreting $\Phi$ as an entropic potential \cite{AgJmr} and
assuming that the volume-dependent Onsager coefficient
$L(\mathrm{v})$ follows a power law of the type
$\mathrm{v}^{\delta}$, where $\delta=1-1/d$, with $d$ the
dimension of the system, one obtains the expression of the rate \cite{JmrAg,AgJmr}
\begin{equation}
J(\mathrm{v}, t)=-\sigma_a \mathrm{v}^{\delta -1}P(\mathrm{v},
t)-D_{a}\mathrm{v}^{\delta}\frac{\partial P(\mathrm{v},
t)}{\partial \mathrm{v}}
\end{equation}
where $\sigma _{a}$ and $ D_a$ are reference constants
\cite{JmrAg}. This expression constitutes the
Louat--Mulheran--Harding ($L-M-H$) law \cite{agreview} proposed
heuristically by assuming that the drift is due to surface tension
effects, considering only the mean curvature. The procedure can be
generalized to the case in which the Gaussian curvature is
important which occurs at small sizes of the grains \cite{AgJmr}.

In the offered modeling we would explore another degree of
freedom which is $R$, the radius of a molecular cluster or a
crystal. It appears naturally in the modeling being of the form
of Fokker-Planck type \cite{AGJS_PSSB}.

\subsection{An example: $2D$ modeling of a crystal growth in complex environment and its MNET features}

A multigrain growth has also been considered as a discretized
process in terms of MNET in both space and time domains
\cite{ausl}. The model is mesoscopic in the sense that each cell
can contain (i) a liquid unit of phase ${\sc l}$, (ii) a crystal
unit, (iii) a small particle embedded in some melt or (iv) a solid
large particle. Initially, each cell of the lattice contains
either a large particle with some probability $P_l$, or a small
particle embedded in some liquid ${\sc l}$ with a probability
$P_s$, or a liquid unit with a probability $1 - {P_l} - {P_s}$.
Nucleation is induced by simultaneously turning a number $n$ of
cells into initial solid units, supposed to be randomly dispersed
in the initial melt. Since the process occurs over long time
scales, thermodynamic (i.e. equilibrium-like) quantities can be
mapped into multigrain growth probability rules, as follows. At
each growth step, all $i$ cells containing some liquid phase, i.e.
${\sc l}$-cells and cells with a small particle, in contact with
mesoscopic cells are selected. The probability $P_i$ to grow the
phase on the cell $i$ is given by a classical thermodynamic
argument as $P_i \sim exp(-{\Delta {G_i}}/{{k_B} T})$, where
$\Delta {G_i}$ is the gain of free energy. Usually, it can be
decomposed into two terms: a bulk contribution depending on the
driving force and a local surface contribution which is
proportional to a chemical bond energy $E_b$, cf. Eq,
(\ref{minimum work}).
%Here you have your
%-> eq '(3)  <-
%to be mentioned

The algorithm on which the discrete modeling is based, combines a
mechanism for pushing and/or trapping of particles, a chemical
reaction, and crystal growth kinetics. The model is an adapted
Eden model \cite{ma2}.
%starting from your
%-> eq (2) Pi ~ exp(-D Gi/kT),
The change in energy enters in the Boltzmann factor, cf. Eq.
(\ref{peq general}). It is considered to be anisotropic and
depending on the number of neighbors. This leads to rugged
surfaces and facets. Moreover, interaction with unreactive, or
reactive particles has been considered as in the theory of
particle trapping or displacement along growing interfaces,
elaborated by Uhlmann et al., and called the $U-C-J$ model
\cite{ucj}. The particles are trapped or pushed by the mobile
solid/liquid interface depending on (i) the surface tensions at
solid/liquid and particle/liquid interfaces, (ii) the particle
(impurity) size and (iii) the growth velocity of the solidifying
front. (A similarity between the $U-C-J$ and what we have proposed
in sec. 4 has to be noticed, cf. Fig. 3 therein.) For a fixed
particle size (see, the FSE effect involved in the modeling
presented in sections 3 and 4), the particles are supposed to be
trapped by the front if the growth velocity of the interface is
higher than a critical value, a physical situation that suits
perfectly our type of modeling. This critical value is controlled
by the particle size and the interfacial tension energies
\cite{ucj}.

Since, while growing the phase in an entropic milieu, the bulk
contribution is roughly constant in the system for isothermal
conditions, only the local surface contribution is needed for
measuring the probability of growth, i.e. ${P_i} \sim
exp(-{g_{nn}}{N_i} )$ where $N_i$ is the number of
nearest-neighboring ($nn$) units belonging to the same grain and
${g_{nn}} = {{E_b}\times {nn}}/{{k_B} T}$. This expresses an
anisotropic locally preferred kinetic growth along the main
directions. For high positive values of $g_{nn}$, square-like
grains are growing \cite{ma3}. For large $g_{nn}$, smooth grain
faces are obtained.

>From a kinetic point of view, it is accepted that the grains grow
anisotropically with a four-fold symmetry along the $a-b$ planes,
the edges of the grains being oriented along the $(10)$ and $(01)$
crystallographic directions. This has been implemented in the
choice of the parameter value $g_{nn}$.

In order to simulate best the chemical reaction process, the
mesoscopic cells for which the nearest and next-nearest neighbors
do not contain any particle are excluded from the growth cell
selection. Indeed, the growth of the phase on these mesoscopic
cells is assumed to be improbable because of the local deficiency
of the basic component around these cells. Taking into account the
probabilities $P_i$ of all possible growing cells, one specific
growth site is randomly selected.

The simulations \cite{ma2,ma3,ausl} put into evidence the effect
of the particle (no matter whether a 'domestic' or testing
particle, or some impurity particle)  size distribution itself on
the resulting microstructure, which is of special interest of the
underlying study. Even though, the MNET-based model considers only
two different particle (impurity) sizes, the model predicts that
the refining of the impurity particles leads to better samples. It
should also be underlined that a discrete MNET-oriented modeling
can be thought of as a useful working extension of many of its
continuous versions \cite{JMR,JmrAg,AgJmr}.

Some extension from $2D$ to $3D$ is rather straightforward but not
too trivial: Just as difficult as going in any simulation of
Eden-type systems when counting sites during spreading must be
done with care, see \cite{ma2}, and refs. therein.

\section{Deterministic part of the approach to a sphere growth
controlled by mass-convective fluctuations: the role of curvatures and
nonequilibrium boundary effects}

In this section, we consider two physically relevant cases of the
boundary condition (BC) prescribed for the growing, e.g. protein
(non--Kossel \footnote{Non-Kossel crystals are defined as complex
structures with several molecules per unit cell in inequivalent
positions \cite{CHERN1}.}), spherical crystal. In the first case,
we propose an essential modification of the curvatures'
contribution to the BC. This includes the Gaussian curvature. This
condition is of equilibrium type. In the second case, we propose
"Goldenfeld-type" condition \cite{Goldenfeld} which is of
nonequilibrium type. In the equilibrium case we would like to
offer a more detailed analysis, whereas in the nonequilibrium
case, because the analytic solution to the problem cannot be
found, we only discuss the asymptotic limit in a brief way.

Let us assume that initially at $t=0$ the growing object is an
ideal sphere of radius $R_0$ and the density of the sphere is $C$.
At time $t>0$ the radius of the growing sphere is equal to
$R=R(t)$. From the mass conservation law \cite{AGJS2} an evolution
equation that gives the speed of the spheroidal crystal formation
arises, and looks as follows
\begin{equation}
    \frac{dR}{dt} = \frac{c_s(1/R)}{C-c_s(1/R)}{v}(R),
    \label{drdt1}
\end{equation}
where $c_s=c_s(1/R)$ is the curvature-dependent surface
concentration, and $v(R)$ is the velocity of incoming macroions,
both taken at distance $R$ from the sphere center. The
concentration $c_s$, prescribed at the boundary, is derived under
the assumption of local thermodynamic equilibrium at the boundary.
It has the form of the well-known linearized Gibbs--Thomson
relation \cite{AGJS2}
\begin{equation}
    c_s = c_0 (1 + 2\Gamma {1\over R}),
    \label{GT1}
\end{equation}
provided that $\Gamma$ is the so-called Gibbs--Thomson or
capillary constant, which is usually of the order of $10 nm$ for
lysozyme crystal \cite{CHERN1}, $c_0$ is an equilibrium
concentration for the planar surface, practically for $R \gg R_0$.
Based on the knowledge of formation of droplets viz (vapor)
condensation processes, one typically defines $\Gamma $ as
\cite{huang}
\begin{equation}
    \Gamma = {{\sigma m}\over {\rho_d k_B T}},
    \label{GTlength}
\end{equation}
where $m$ is the mass of the vapor-phase atom (unit) and $\rho_d $
stands for the density of the droplet's material, being here
identified with $C$ of Eq. (\ref{drdt1}); $\sigma $ and $T$ have
their usual meaning (see above).
The above definition can also be adopted to our case \cite{vekir}. \\
The solution of Eq. (\ref{drdt1}) with Gibbs--Thomson boundary
condition, Eq. (\ref{GT1}), for ${v}(R) = v_{mi} = const.$ reads
\cite{AGJS2}
\begin{equation}
    R-R_0-(R_C+2\Gamma)ln \frac{R+2\Gamma}{R_0+2\Gamma}=\sigma_0 v_{mi} t,
    \label{sol1}
\end{equation}
and its large time asymptotic solution becomes
\begin{equation}
    R\sim t.
    \label{sol1_1}
\end{equation}

\noindent Here
\begin{equation}
 \sigma_0 = {c_0 \over {C-c_0}} ,
    \label{sigma0}
\end{equation}
and
\begin{equation}
R_c=2\sigma_0 \Gamma
\label{R_c}
\end{equation}
is a critical nucleus' radius. $\sigma_0$ is an equivalent of the
bulk
supersaturation%
\footnote{
    Except that the original Gibbs--Thomson condition involves a firm curvature-dependent
contribution, see Eq.~(\ref{GT1}),
    it is rather hardly applicable to crystallization under high bulk
    supersaturation. This fact imposes some limits to the magnitude of the overall chemical-potential based driving force of crystal growth as a whole~\cite{CHERN1,VekilovAlexander}.
}.

\begin{figure}
\begin{center}
  \includegraphics[width=0.8\textwidth]{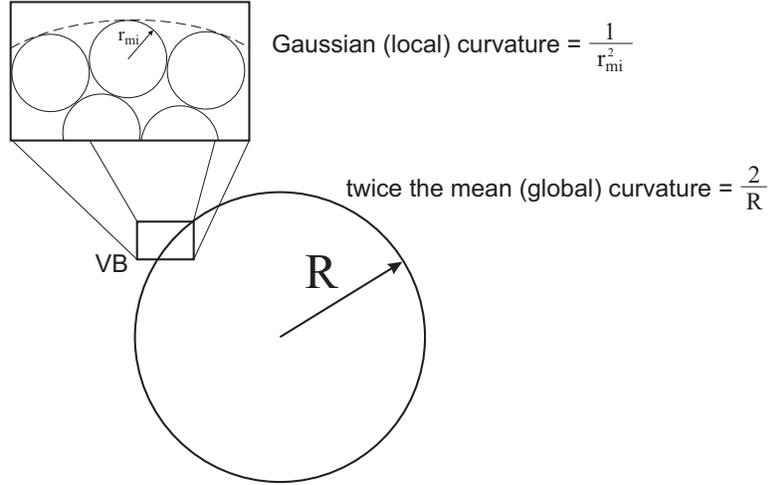}\\
  \caption{Schematic picture of the mean (global) and Gaussian (local) curvature problem.
  Twice the mean curvature of the growing sphere of radius $R$
  equals $2\over R$ and is a global curvature measure. Upon
  magnifying the visualisation box (VB) a local picture appears,
  at a given time instant $t$ and packing conditions $\eta$ (see caption to Figure \ref{rt}), for which the Gaussian curvature of
  the building block viz the spherical molecule of radius $r_{mi}$
  reads $1\over r_{mi}^2$. The overall curvature contribution to
  the equilibrium BC of Gibbs--Thomson type, Eq. (\ref{GT1}),
  reads $\delta_T^2\over {r_{mi}R}$, $\delta_T$ - Tolman parameter
  \cite{tolman}, and leads to a modified BC, Eq. (\ref{GT1}), of
  importance for crystal composed of finite-size (or non-point
  like) building blocks.}
\label{curvature}
\end{center}
\end{figure}

As the first essential modification (equilibrium type) of the
curvature contribution to the BC we propose to include the
Gaussian curvature. Schematic picture of the mean (global) and
Gaussian (local) curvature problem is presented in Figure
\ref{curvature}. This correction can be important, because of
large linear dimension of the growth units in comparison with
linear dimension of the other solution components, so the local
curvature can play important role, especially when we deal with
macromolecular islands built up of a few units. Now the BC takes
the form
\begin{equation}
    c_s = c_0 (1 + 2\Gamma {1\over R} + {\delta_T^2\over r_{mi}} {1\over R})=c_0(1+2\tilde{\Gamma}\frac{1}{R}),
    \label{GT2}
\end{equation}
where
\begin{equation}\label{Gamatylda}
    \tilde{\Gamma}=\Gamma+\frac{\delta_T^2}{2r_{mi}}
\end{equation}
and $1/(r_{mi}R)$ is a Gaussian curvature, $r_{mi}$ is the radius
of the crystal building unit (for lysozyme protein $r_{mi}\approx
1.5 nm$) and $\delta_T$ is a Tolman length.  $\delta_T$ is defined
as the difference between the radius of the surface of tension and
the radius of the equimolar dividing surface (for lysozyme protein
$\delta_T\approx 3.5 nm$) and depends on the packing coefficient
of the crystal. The Tolman length can be related to the
superficial density at the surface of tension \cite{tolman}. Note
that by postulating $c_s$ in the form of Eq. (\ref{GT2}), with
$\delta_T$ involved, we somehow induce, for a given time instant
$t$, elastic effects at the crystal boundary. This is because
$\delta_T$ is also a measure of rigidity for bending of a curved
piece of the crystal boundary, a physical fact well known to those
studying biomembrane formation \cite{Blockhuis}. Thus, we may,
also by means of having $\delta_T$ involved in the BC, arrive at
elastic contribution to the crystal growth \cite{AGJS_PSSB}: This
situation resembles a well-known fact that an improperly placed
boundary molecule, or some impurity, exerts an additional
strain-stress
field in its very vicinity \cite{VekilovAlexander}.\\
The solution of Eq. (\ref{drdt1}) with BC prescribed by Eq.
(\ref{GT2}), with (\ref{Gamatylda}) is of the same type as the one
given by Eq. (\ref{sol1}) but now
\begin{equation}\label{Rctylda}
    R_c=\tilde{R_c}=2\sigma_0\tilde{\Gamma}
\end{equation}
has to be inserted, cf. Eq. (\ref{R_c}).

\begin{figure}
\begin{center}
  \includegraphics[width=0.9\textwidth]{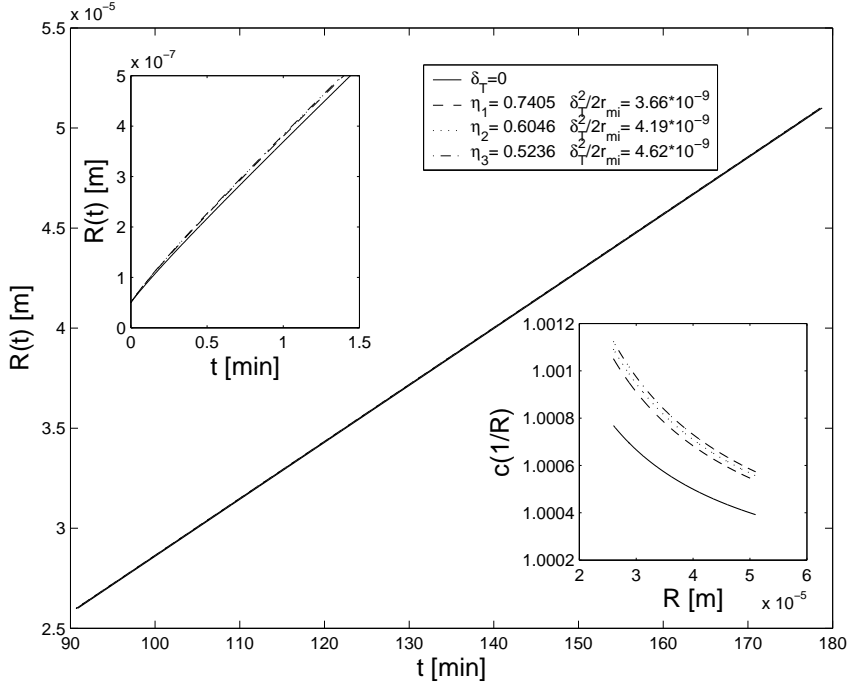}\\
  \caption{The time dependence of the crystal radius $R(t)$ for original (\ref{GT1})
  and modified (\ref{GT2}) BCs of Gibbs-Thomson (and Tolman) types taken for three different crystal structures
  (cubic close packing $\eta_1$,
  hexagonal lattice $\eta_2$ and cubic lattice $\eta_3$). The
lower-right inset shows differences in behavior of the surface
concentration with and without Gaussian curvature (finite-size or
Tolman type) correction to the surface concentration, while the
upper-left inset points to early-stage differences in tempo of
crystal formation. These differences stand also over longer time
period but they cannot be seen as separated curves for time
interval chosen to fit our model to experimental data ($C = 500$
[arbitrary units], $c_0 = 1$ [arbitrary units],
  $\Gamma = 1 \cdot 10^{-8} m$, $R_c=4\cdot 10^{-11} m$, $R_0 = 5\cdot 10^{-8} m$, $v_{mi} = 2.3 \cdot 10^{-6} m/s$).}
\label{rt}
\end{center}
\end{figure}

The dependence of $R(t)$ is presented in Figure \ref{rt}. The
lower-right inset to Figure \ref{rt} shows differences in behavior
of the surface concentration with and without Gaussian curvature
correction to the surface concentration, while the upper-left
inset points to early-stage differences in tempo of crystal
formation expressed in terms of $R$ vs $t$ dependence. These
differences stand also over longer time period but they cannot be
seen as separated curves for time interval ($1.5$ hour) chosen to
fit our model to experimental data \cite{chowAPL}. We can see that
Gaussian curvature influences (speeds up) the growth rate just by
increasing the local surface concentration, especially for the
early stage of the growth process when $R/r_{mi}$ is relatively
small.

Our results adjust well to the experiment. Chow et al.
 \cite{chowAPL} observed constant growth rate of the lysozyme
spherulites, what is characteristic when the growth is controlled
by surface kinetics rather than by volume diffusion. In the
experiment the crystal density is 25 times bigger than lysozyme
concentration in the solution. In our calculation supersaturation
is equal 500. This discrepancy implies that the narrow interfacial
region (double layer) designed for our model is definitely a
protein-poor region.

In the second case we propose "Goldenfeld-type" correction
\cite{Goldenfeld}. Now we assume that the surface is away from
local equilibrium and deviation from this state is proportional to
the growth velocity of the interface
\begin{equation}
    c_s = c_0 (1 + 2\tilde{\Gamma}\frac{1}{R} - \beta_G \frac{d R}{d t}),
    \label{goldenfeld}
\end{equation}
where $\beta_G$ is a positive kinetic coefficient. Solving Eq.
(\ref{drdt1}) with nonequilibrium boundary condition, Eq.
(\ref{goldenfeld}), one gets for long times the linear solution of exactly
the same type than that given by
(\ref{sol1_1})
%\begin{equation}
%    R\sim t,
%    \label{sol_goldenfeld}
%\end{equation}
which can reflect the asymptotic temporal behavior of spherulites.
It can be shown that when $\beta_G$ increases the growth becomes
slower. At early stage of the growth process the influence of
$\beta_G$ on $R(t)$ is strongly nonlinear. For long times, the
growth is linear in time and $\beta_G$ influences only the growth
rate (slope of the curve, cf. Figure \ref{rt}) \cite{AGJL_IJQCh}.

Looking at the Eq. (\ref{drdt1}) we can see that the growth
process depends not only on the supersaturation of the solution.
It turns out to be very important that the macroion velocity,
which depends on the electrostatic interactions of the macroion
with the solution (we have to mention that the solution is an
electrolyte), also depends on the viscosity of the solution, and
moreover, on the size of the growth units. It is sometimes
possible that the growth unit will be made of few macromolecules,
so that in the viscoelastic complex environment that we actually
describe, the aggregates (made up of $2-5$ macromolecules) will
have different velocities, especially when compared with that of
the single macroion because the diffusion coefficient can now be
aggregate-size dependent \cite{coniglio}. This experimentally
justified observation \cite{wilcox} has its profound impact on the
so-called intrinsic properties of the viscoelastic solution
\cite{plonka}, making them time-sensitive, thus imposing an
intrinsic time scale, presumably powerly "correlated" with the
observation time scale \cite{BBMandel}. It will be demonstrated in
the subsequent section.

\section{Sphere growth controlled by convective fluctuations: the role of
time-dependent viscosity and finite-size effects}

In accordance with the considerations of the previous section,
here we will complement the deterministic description of the
kinetics of crystal growth by incorporating thermal fluctuations
of the bath through fluctuations of the velocity of the incoming
macroions (or growth units) \cite{LuczkaNiemec02}, and relating
them with the finite size of the particles at the locally highly
concentrated regime. In this regime, the viscoelastic properties
of the solution become important and must also be incorporated
into a consistent description.

\begin{figure}
\begin{center}
  \includegraphics[width=0.9\textwidth]{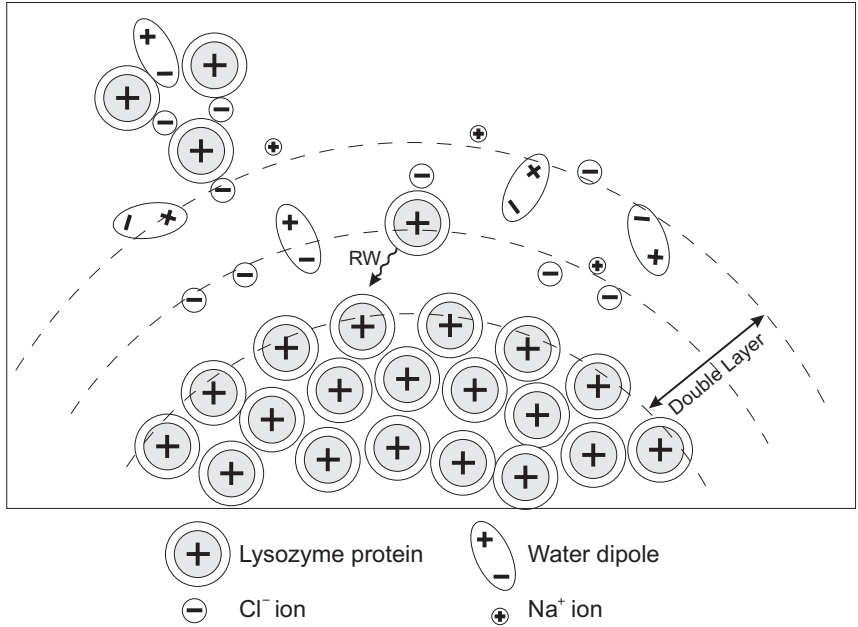}\\
  \caption{Naive picture of the vicinity of the crystal surface.
  Building unit (positively charged lysozyme protein), which is surrounded by
  neutral (water dipoles) and oppositely charged particle (counterions),
  performs random walk (RW) with a constant velocity in an electrostatic double layer.
  We can see that the crystal surface is not smooth and the local (Gaussian)
  curvatures are shown (see caption to Figure \ref{curvature}).
  In the left top corner a triplet has already been formed.}
  \label{dl}
\end{center}
\end{figure}

To perform this description, we will adopt an effective medium
theory \cite{S-EJPCM} in which the diffusion of a "test" macroion
through a saturated solution can be analyzed in terms of the
single-particle distribution function $f(\vec{r},\vec{u},t)$,
depending on the position $\vec{r}$ and instantaneous velocity
$\vec{u}$ of the macroion. In accordance with mesoscopic
nonequilibrium thermodynamics formalism, in this case the entropy
per unit volume, $s(t)$, can be expressed in terms of the Gibbs
entropy postulate  \cite{S-EJPCM,gradtemp}
\begin{equation}
s(t)=s_{le}-k_{B}\int f \ln \frac{z}{z_{le}}d\vec{u}, \label{p.
gibbs}
\end{equation}%
where  $s_{le}$ is the entropy at local equilibrium, cf. Eq.
(\ref{entropy postulate}), and we have introduced the fugacity
$z=\alpha\,f(\vec{r},\vec{u},t)$, with $\alpha$ the activity
coefficient characterizing the interactions of the macroion.
$z_{le}$ is the fugacity at local equilibrium. The activity
coefficient $\alpha$ will contain two contributions since the test
macroion interacts with both, the solution of macroions
$\alpha_{s}$ (mostly, in the vicinity of the crystal boundary) and
with the crystal $\alpha_{c}$, see Figure \ref{dl}. The activity
of the solution can be expressed as $\alpha_{s} = e^{c^{-1}p/k_B
T}$, with $c^{-1}p$ the compressibility factor in which $p$ is the
excess of osmotic pressure \cite{S-EJPCM} and the concentration
field is defined by $c(\vec{r},t)=\int
f(\vec{r},\vec{u},t)d\vec{u}$. The explicit expression of
$c^{-1}p$ can be obtained in the context of different theories on
electrolyte solutions \cite{hillBOOK}. The crystal activity
$\alpha_{c}$ can, in general, contain effects related with
entropic, energetic or geometrical constrains, as indicated in
section 2, \cite{AGJS_PSSB}.

Following the procedure of MNET already outlined, one may show
that the Fokker-Planck equation describing the evolution of
$f(\vec{r},\vec{u},t)$ is \cite{S-EJPCM,SteadyState}
\begin{equation}  \label{GFPE}
\frac{\partial f}{\partial t}+\nabla \cdot \left( \vec{u}f\right)
-k_{B}T \nabla\left[\ln \alpha_{c} + \ln \alpha_{s} \right] \cdot
\frac{\partial f}{\partial \vec{u}} =\frac{\partial }{\partial
\vec{u}}\cdot \beta(t) \left[ \vec{u}f+\frac{k_{B}T}{m}
\frac{\partial f}{\partial \vec{u}}\right],
\end{equation}
where the third term on the left-hand side of the equation
contains the forces acting on the macroion. The right-hand side of
this equation contains the contributions of Brownian motion of the
macroion in which the time-dependent friction coefficient
$\beta(t)$ introduces memory effects \cite{adelman,nonmarkov}. As
mentioned in the previous section, these effects are related to
the viscoelastic properties of the solution (time-dependent
viscosity), and modify the intrinsic time scale of the particle
dynamics.

Several models for the friction have been used in the literature
of model crystal growth \cite{LuczkaNiemec02}. However, since we
are interested in analyzing the effects of the finite size and
inertia of the macroions, then we must take into account that in
this case a general expression for the friction coefficient is
\cite{SteadyState,S-EJPCM}: $\beta^{-1}(t) \propto {\cal
L}^{-1}\left[1+ \left(\tau_{D} \omega \right)^{\gamma_c}
\right]^{-1}$, with the symbol ${\cal L}^{-1}$ denoting the
inverse Laplace transform, $\omega$ the frequency and
$\tau_{D}=\frac{r^2_{mi}}{D_0}$ a characteristic diffusion time,
with $D_0=\frac{k_B T}{6\pi \eta_{id} r_{mi}}$ the diffusion
coefficient of the macroion at infinite dilution, $r_{mi}$ is the
macroion radius and $\eta_{id}$ the viscosity of the solvent at
infinite dilution. In this relation, the exponent $\gamma_c$ may
in general be a function of the macroion concentration at the bulk
\cite{S-EJPCM}. In order to make an analytical progress, for
simplicity one may use the expression of $\beta(t)$ obtained in
the context of the generalized Fax\'{e}n theorem
\cite{SteadyState,mazur-bedo}
\begin{equation}
\beta^{-1}(t)=\beta^{-1}_0
\left[\sqrt{\frac{\tau_{D}}{t}}-exp\Bigg({\frac{t}{\tau_{D}}}\Bigg)
Erfc \left(\sqrt {\frac{t}{\tau_{D}}}\right)\right], \label{beta}
\end{equation}
where $\beta_0=6\pi \eta_{id} r_{mi}$ is the Stokes friction
coefficient. Notice that Eq. (\ref{beta}) is the inverse Laplace
transform of $\beta^{-1}(\omega)=\beta^{-1}_0\left[1+(\tau_{D}
\omega)^{1\over 2}\right]^{-1}$, and that at times $t<<\tau_{D}$,
it reduces to $\beta^{-1}_0 \sqrt{\frac{\tau_{D}}{t}}$.

To obtain the velocity $\vec{v}(t)$ of the macroions entering into
the growth rule (\ref{drdt1}) in which, however, the macroion
velocity has been assumed constant, see legend to Figure \ref{rt},
one may use Eq. (\ref{GFPE}) to obtain the hierarchy of evolution
equations for the moments of $f$, and from it construct the
equation \cite{S-EJPCM}
\begin{equation}  \label{Eq.Diff}
\frac{\partial c}{\partial t}= -k_{B}T\nabla\cdot \left[\beta^{-1}
c\nabla\ln \alpha_{c}\right]+ \nabla \cdot \left[ D_{c} \nabla
c\right].
\end{equation}
which describes the evolution of the macroion in the diffusion
regime. Here, we have defined the collective diffusion coefficient
$D_c \equiv k_{B}T\beta^{-1}\left(1 + \frac{\partial \ln
\alpha_s}{\partial \ln c} \right)$. Particular expressions for
this coefficient can be modelled within the scope of, for example,
Kirwood theory of solutions. In a mean field approximation, the
pressure $p$ can be approximated by an expansion of the form $p k_B T\left[c - B(T) c^2\right]$, where $B(T)$ is a virial-type
 coefficient incorporating the specificities of the pairwise interactions
among macroions. In general, $B(T)$ may depend on the radial
distribution function, and then on spatial correlations.
Substituting this expansion into $\alpha_{s} = e^{c^{-1}p/k_B T}$,
taking the logarithm and the corresponding derivative of the
definition of $D_c$, one obtains the time-dependent diffusion
coefficient $D_c = k_{B}T\left(1 -B(T)\right) \beta^{-1}(t)$.It is
interesting to notice that $T\left(1 -B(T)\right)$, resembling an
"effective" temperature \cite{S-EJPCM}, implies that macroion
diffusion has a nonlinear dependence on $T$, which could be
relevant in the crystal structure, as was already mentioned. From
Eq. (\ref{Eq.Diff}), it follows that in the diffusion regime the
velocity of the macroions will satisfy a Langevin equation of the
form
\begin{equation}  \label{v(t)}
\frac{d r(t)}{d t}=v(t) = \beta^{-1}(t) \left[F_d(t) +
F^R(t)\right],
\end{equation}
where for simplicity we have considered the unidimensional case
and $F^R(t)$ constitutes a random force due to thermal
fluctuations. Moreover, we have defined the deterministic force
$F_d(t)=\int \left[k_{B}T\frac{d \ln \alpha_{c}}{d
r}\right]c(r,t)dr$, due to the interaction of the macroion with
the crystal. Notice that the fluctuating part of the velocity
satisfies the usual conditions of zero mean and $\langle
v(t)v(t_1)\rangle= \beta^{-1}(t-t_1)$ \cite{adelman}.

In a first approximation, one may assume that in the vicinity
(double layer) of the crystal surface, the electrostatic force
(attracting macroions) is constant, $F_d=F_0$, \cite{AGJS2}. This
assumption implies that $\alpha_c$ depends upon the minimum
reversible work necessary to move the macroion a certain distance
$\Delta r$, $\alpha_c =e^{F_0\Delta r/k_BT}$, see Eq.
(\ref{minimum work}). On taking into account the above
assumptions, near to the crystal, the growth rule (\ref{drdt1})
can be written in the approximated form

\begin{equation}  \label{Growth-stochastic}
\frac{d R(t)}{dt} = \sigma_{R} \beta^{-1}_0
\sqrt{\frac{\tau_D}{t}} \left[F_0+F^R(t)\right]
\end{equation}
where $\sigma_R=\frac{c_s}{C-c_s}$. In the present approximation,
$\sigma_R$ can, for instance, be expressed in the form

\begin{equation}  \label{sigma_R}
\sigma_R=\sigma_0 \frac{1+{2\tilde{\Gamma}\over
R}}{1-{2\tilde{R}_c\over R}},
\end{equation}
after using Eqs. (\ref{GT2}), (\ref{Gamatylda}) and
(\ref{Rctylda}). Note that for $t >> 1 $ $\sigma _R \to \sigma _0$
holds. Moreover, notice that for $c_s$ we have to take here,
depending on what we wish to model, either $c_s$ from (\ref{GT2})
(equilibrium BC: non-Kossel crystals) or the one from
(\ref{goldenfeld}) (nonequilibrium BC: biopolymeric spherulites).
Especially, the latter seems to be a challenge for extensive
numerical modeling, which is, however, left for a future task. In
both cases mentioned, however, the FSE account is thoroughly
manifested, cf. Figs 1-4.

\begin{figure}
\begin{center}
  \includegraphics[width=0.9\textwidth]{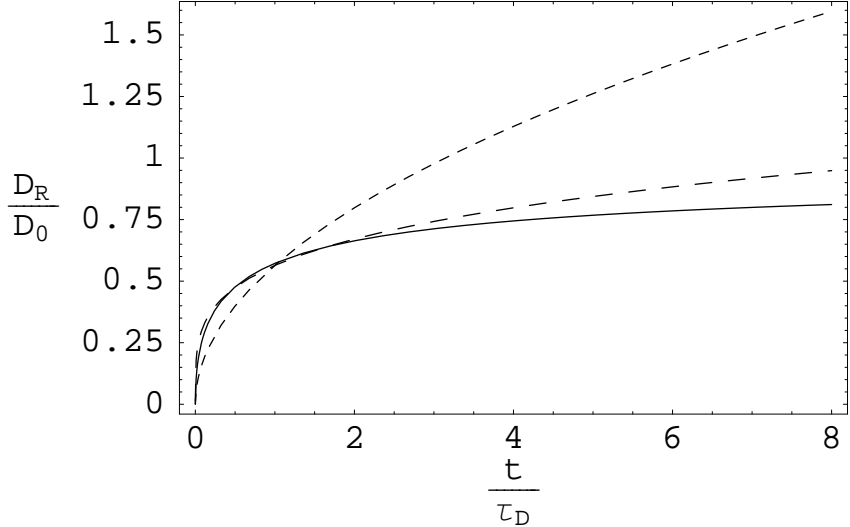}\\
  \caption{Reduced diffusion coefficient $\frac{D_R}{D_0}$ as a function of
the reduced time $\frac{t}{\tau_D}$, for $D_0 \simeq 10^{-6}cm^2
s^{-1}$ and $\tau_D \simeq 10^{-8} s$. The solid line represents
the integral of Eq. (\ref{beta}), whereas the short-dashed line is
represented by Eq. (\ref{Dyf_R}). In the deterministic case, the
integration of Eq. (\ref{Growth-stochastic}) can be compared with
that of Eq. (\ref{sol1}) with the parameters $\tilde{\Gamma}$ and
$\tilde{R}_c$, from which follows that at sufficiently large times
the left-hand side behaves as $R$, whereas the right hand side
will be proportional to Eq. (\ref{D_R}). Since the solid line
tends to a constant asymptotic value, then from the figure it
follows that for sufficiently long times, the radius of the
crystal tends to a maximum value. In contrast, $\beta^{-1}_0 \sqrt{\frac{\tau_{D}}{t}}$ implies a "pure" superdiffusive growth.
The long-dashed line, corresponding to a $t^{\frac{1}{4}}$
behavior, has been included for comparison. It can be obtained
form $\beta^{-1}(t) \propto {\cal L}^{-1}\left[1+ \left(\tau_{D}
\omega \right)^{\gamma_c} \right]^{-1}$, when the exponent
$\gamma_c = 1/4$.}
  \label{diffusion}
\end{center}
\end{figure}

\begin{figure}
\begin{center}
  \includegraphics[width=0.9\textwidth]{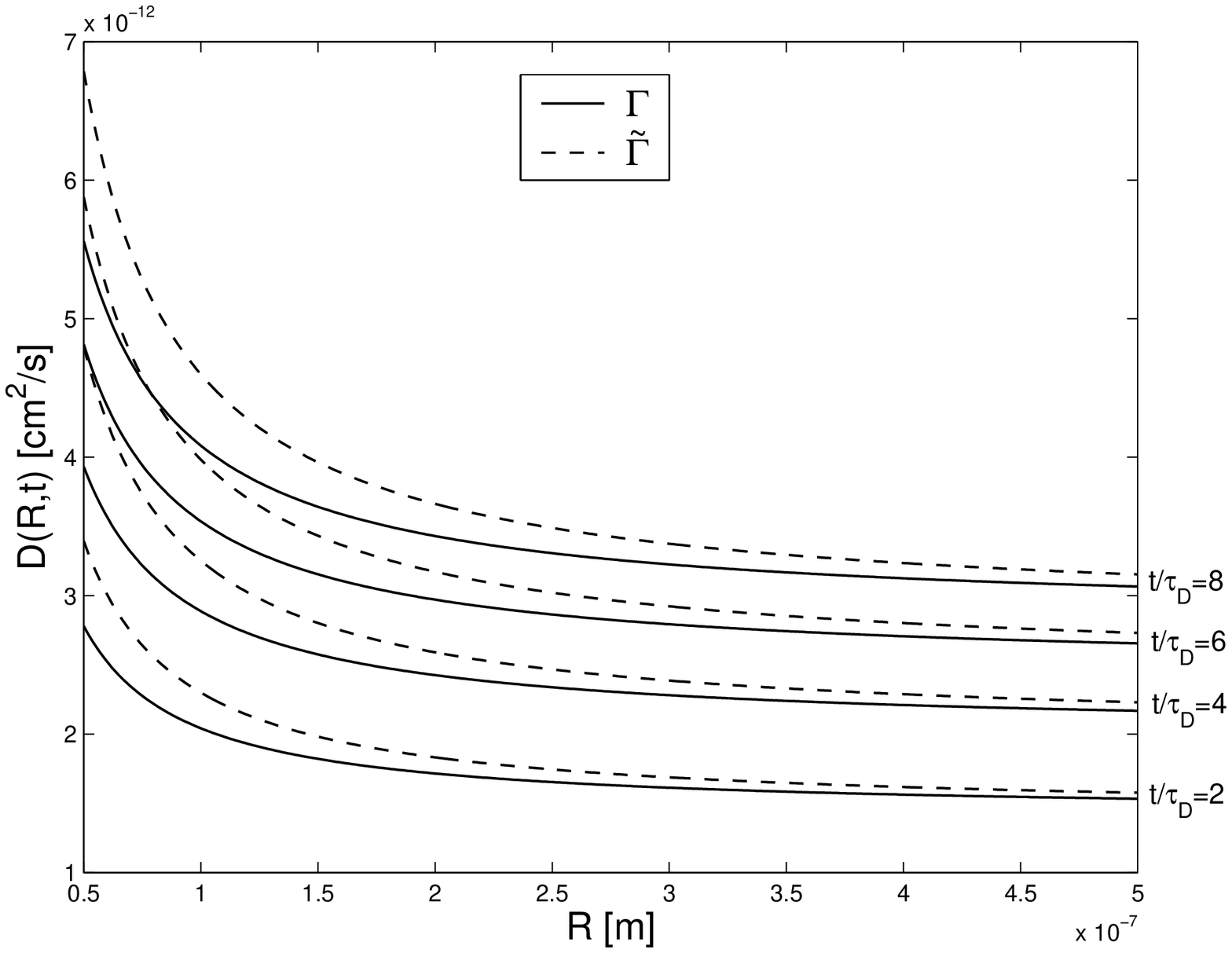}\\
  \caption{Mean course of the overall diffusion-type behavior of the
  non-Kossel (model) crystal growth.
  The curvature effect and FSE are largely pronounced
  for $\gamma_c={1\over 2}$ and some consecutive time steps $t$.
  The same type of temporal behavior has been proposed in \cite{LuczkaNiemec02}
  for algebraic correlations with exponent $\gamma_c={1\over 2}$.
  Solid lines are plotted on the basis of Eq. (\ref{Diffcoeff}),
  whereas the dashed lines on its equivalent with $\tilde{R_c}=R_c$ and
  $\tilde{\Gamma}=\Gamma$ (for details see caption to Figure \ref{diffusion}).}
  \label{dyf}
\end{center}
\end{figure}

Now, in order to show the relation existing between the present
approach with that of Sec. 2,  for simplicity we will consider the
case in which the deterministic part $F_0$ can be neglected. As a
consequence, Eq. (\ref{Growth-stochastic}) becomes a stochastic
equation with multiplicative noise which is equivalent to the
Fokker-Planck equation \cite{LuczkaNiemec02}
\begin{equation}  \label{FPE-R}
\frac{\partial P(R,t)}{\partial t} =\frac{\partial }{\partial R}
\left[D(R,t)\frac{\partial P(R,t)}{\partial
R}+\frac{D(R,t)}{k_{B}T}  \frac{\partial \Phi}{\partial R} P(R,t)
\right],
\end{equation}
where $P(R,t)$ is the probability density in the $R$-space with
$R$ standing for the radius of the spherical crystal. Moreover, we
have defined\footnote{In another paper \cite{AGJS_PSSB} we have
defined $\Phi=k_BT ln\sigma_R$, i.e. in a Boltzmann-like form. It
automatically implies that the time-dependent part of Eq.
(\ref{Diffcoeff}), $D_R(t)$, has to be defined without $k_BT$ as a
prefactor absorbed in $D_0$. It, however, destroys an
Stokes-Einstein type form of $D_0$, which is not the case of the
present paper in which this form is used explicitely, cf. caption
to Figure \ref{dyf}.} the potential $\Phi \equiv \ln \sigma_{R}$
and the diffusion coefficient
\begin{equation}  \label{Diffcoeff}
D(R,t) \equiv D_R(t) \sigma_{R}^{2},
\end{equation}
where
\begin{equation}  \label{Dyf_R}
D_R(t)= D_0 \sqrt{\frac{ t}{\tau_D}}
\end{equation}
is the time-dependent diffusion coefficient of the incoming
macroions related with $\beta^{-1}$ through \cite{LuczkaNiemec02}
\begin{equation}  \label{D_R}
D_R(t) = k_B T \int^{t}_0 \beta^{-1}(t_1) dt_1.
\end{equation}
Figure \ref{diffusion}, shows the time-dependent part of the
reduced diffusion coefficient as a function of the reduced time.
The solid line represents the integral of Eq. (\ref{beta}),
whereas the short-dashed line is represented by Eq. (\ref{Dyf_R}),
see Eq. (\ref{Diffcoeff}). The long-dashed line represents a
$t^{\frac{1}{4}}$ behavior for comparison. From the solid line it
follows that at short times the crystal grows with a
superdiffusive behavior whereas at larger times, $D_R$ tends to a
constant value, implying that the crystal radius tends to a
maximum value, i.e. to a final equilibrium state. It coincides
well with both the deterministic view of the process, and above
all, with its experimental realizations
\cite{vekir,VekilovAlexander}. It is interesting to realize that
formally both the growth speed and the constrained motion of a
'representative' macroion, feeding the crystal go in a
superdiffusive way. Thus, it seems that both sub-processes
manifest a kind of synchronous temporal behavior.

Finally, notice that from equation (\ref{FPE-R}) it follows that
the deterministic part of the current is
\begin{equation}  \label{}
J(R, t)=-\frac{D(R,t)}{k_{B}T}\frac{1}{\sigma_{R}}\frac{\partial
\sigma_{R}}{\partial R}.
\end{equation}
Comparing this expression with Eq. (\ref{fug2}), one may conclude
that $\sigma_{R}$ plays the role of a fugacity in $R$-space
whereas $\Phi$ the role of a chemical potential containing
geometric restrictions due to the boundary conditions, as
discussed in the previous section \cite{JmrAg}.

\section{Summary and conclusions}

In this paper, we have proposed an integrated static-dynamic
picture of model (spheroidal) crystal growth in a complex milieu,
preferentially of electrolytic type. The complex milieu we have in
mind here could, in general, be a two-component system, containing
some impurities and/or (charged) additives. The spectrum of
examples can range from metallic polycrystals ($L-M-H$ model, sec.
2.1 and 2.2) via superconductors\footnote{$YBCO$-like
superconducting ceramics  grown near a peritectic point i.e., when
a solid is in contact with a liquid and there is incomplete
reaction \cite{ucj}} ($U-C-J$, sec. 2.3) until the soft-condensed
matter objects that are: non-Kossel crystals and (bio)polymeric
spherulites. In fact, the latter suits best our type of modeling
since it really manifests the FSE property due to pronounced sizes
of the macromolecules constituting the (dis)ordered macromolecular
cluster. The goal of readily incorporating the FSE property into
the growth rules governing spheroidal crystal formation has been
successfully achieved at both static as well as dynamic level of
description - therefore our integrated model is called
static-and-dynamic. At the static level of description the goal
has been achieved by proposing a modification of the well-known
Gibbs-Thomson (equilibrium) formula or by additionally including
its nonequilibrium modification that we attributed to Goldenfeld,
who likely did it for the first time almost twenty years ago but
for a diffusive field feeding the growing object
\cite{Goldenfeld}. In this study, however, we have concentrated on
the equilibrium BC with the Tolman-type correction proposed,
emphasizing a pivotal role of the Gaussian curvature in the
realistic view of the process that we have offered. Therefore, the
nonequlibrium type BC is dealt with in a rather sketchy or
qualitative way, but nevertheless, such a possible modification is
worth mentioning, at least for some interesting future task. Some
undoubtful integrity of the picture, or its quite comprehensive
character, at which we have finally arrived, can by no means be
accomplished when no complementary stochastic part had a chance to
appear. This is, in our opinion, the most worth-emphasizing part
of the description, very responsible for unveiling its dynamic
aspects. It could also be worth-presenting when forseeing a
necessity to go beyond some analytic modeling
\cite{S-EJPCM,AgJmr,JmrAg,LuczkaNiemec02}, and to propose
thoroughly computer simulations, carried out in a complementary
way \cite{ausl,ma2,ma3}. Thus, the integrated static-dynamic
picture of the process is offered in sec. 4. It tells us that: (1)
the role of curvature(s) is important; (2) the role of FSE is
equally important, and anticipates the approach to be quite
realistic; (3) the role of velocity fluctuations in time appears
to be crucial in setting up properly the picture: a basic result
of sec. 4 points to the constrained motion of the macroion to be
superdiffusive, with the exponent $1/2$ for an unidimensional
approximation offered therein. Thus, and in particular, a specific
goal to propose a firm mathematical form for algebraic
correlations, somehow underscored in another study
\cite{LuczkaNiemec02} has been achieved too. The account of
correlations in space is not explicitely taken into account in
this study. Implicitely, however, it is involved in it by
proposing Eq. (\ref{Eq.Diff}) with the collective measure, $D_c$,
from which further a Langevin-type as well as Fokker--Planck type
descriptions for model crystal growth (within the integrated
picture) arise. What above all comes out from the proposed
integrated picture, however, is a clear dynamic measure of this
integrity, represented by $D(R,t)$ of Eq. (\ref{Diffcoeff}). It is
seen in a picturesque way in Fig. 5 which tells us that the
process goes naturally in the way that for its final (mature)
growing stages the overall diffusivity of the system, due to the
above mentioned physical factors (FSE, curvature(s), fluctuations)
slows down, this way likely promoting order against disorder in
the complex entropic milieu from which the object emerges. Also, a
difference between Gaussian and non-Gaussian curvature effects on
the diffusive behavior of the system clearly pronounce in the
course of time. Finally, let us underline a remarkable feasibility
of the presented type of modeling to serve for fitting to some
experimental curves, or to relate the proposed theory to the
experiment. An example of such an ability has been offered by Fig.
2, i.e. the case of biopolymeric spherulites, for the kinetic
coefficient, ${\beta _G}\to 0$, i.e. when the nonequilibrium BC
gets finally equilibrated. Other fits can be done for non-Kossel
crystals as well \cite{VekilovAlexander,vekir}. Last but not
least, let us juxtapose some features of MNET, because this
thermodynamic-kinetic formalism enabled to offer such an
integrated view of the process. These are
\cite{JMR,JMR1,mazur-bedo,nonmarkov}:
\begin{itemize}
\item MNET provides a description of kinetic proceses taking place
at the mesoscale (and involving mesostructures) in which
fluctuations are important. \item The theory provides in general
expressions for the rates (nucleation, growth, reaction, etc.)
which are in general nonlinear functions of the driving forces.
They are derived in very general conditions even when
mesostructures are embedded in an inhomogeneous bath. \item The
theory provides kinetic equations of the Fokker--Planck type for
the evolution of the probability distribution. From them one can
derive expressions for the moments of the distribution which are
related to physical quantities which can be compared with
experiments.
\end{itemize}

\section*{Acknowledgements}
The authors dedicate this paper to Prof. Andrzej Fuli\'nski on the
occasion of his 70th birthday. A support by 2P03B 03225
(2003-2006) is to be mentioned by A.G. and J.S. This work has been
partially supported through Action de Recherche Concert\'ee
Programs of the University of Li$\grave{e}ge$ ARC 94-99/174 and
ARC 02/07-293. M.A. thanks also RW.0114881-VESUVE program for
other partial support. I.S.H. acknowledges UNAM-DGAPA for economic
support. A.G. thanks the Organizers  of the $17$th Marian
Smoluchowski Symposium on Statistical Physics for inviting him to
present a lecture the material of which has partly been contained
in the present paper.\\

% \newpage

\newpage

%{\Large Figure Captions} \vspace{1cm} \indent

%Fig. 1 \\

%.\\

%Fig. 2 \\

%.\\

%Fig. 3 \\

\end{document}